# Source-Independent Fault Detection Method for Transmission Lines in IBR-Dominated Grids

Julio Rodriguez, *Student Member, IEEE,* Isaac Kofi Otchere, *Student Member, IEEE,* and
Reza Jalilzadeh Hamidi, *Senior Member, IEEE*

*Abstract*—This paper proposes a source-independent method for the detection and classification of faults along Transmission Lines (TLs). It aims to reduce the protection issues arising from Inverter-Based Resources (IBRs). Inspired by Power Line Communication (PLC), the proposed method utilizes high-frequency carrier waves which are sent from either side of a TL over each phase. As faults disrupt the propagation of carriers, the receiving carrier waves before and during faults exhibit differences. Based on this principle, the proposed method continuously compares the receiving carrier waves with a short history of them to detect and classify faults.

The performance of the proposed method was evaluated using EMTP-RV and MATLAB, and compared to traditional phasor-based distance relays. The simulation results confirm the capability of the proposed method in detection and classification of different faults regardless of power sources types.

## I. Introduction

THE quest for new power resources and advancement in technology have led to the integration of Inverter-Based Resources (IBRs) into power grids. Although IBRs offer several preferable functionalities [1], heavy IBR contamination negatively impacts grids' protection systems as the fault response of IBRs and electromechanical generators are significantly different [2]. The drastic difference between the fault response of IBRs and conventional generators makes phasor-based relays inefficient in detecting faults in highly IBR-contaminated grids [3]. Therefore, several methods for overcoming the protection challenges arising from IBRs have been suggested in the literature.

One of the solutions proposed in the literature is to employ adaptive protection schemes which largely rely on communications to dynamically adjust the settings of relays in accordance with the grid's status. Despite the advantages of adaptive schemes in detection of faults in IBR-based grids [3], their dependency on communications makes them vulnerable to communication failures and cyber-attacks [4], [5].

Another remedy is the use of Traveling Waves (TWs) induced by faults. However, the performance of TW-based protection is highly dependent on the grid's and fault's parameters [6]. Thus, TW-based protection has to be backed up with other protection schemes [7]. Another solution is to incorporate control features into IBRs to mimic electromechanical generators' fault responses to assist phasor relays in fault detection [8]. Although this approach is quite successful in several cases, a large portion of IBRs cannot generate high fault currents as electromechanical generators produce since their prime energy sources are insufficient (e.g., solar panels) [9], [10]. For solving this issue, the deployment of energy storage devices (e.g., supercapacitors) is suggested in [5], [11]. However, the associated costs are considerable. Another issue with incorporation of controllers in IBRs to mimic generators' fault responses is that the controller should be made aware of the occurrence of faults. One way to inform the controller of faults is to use voltage drops at IBR terminals as a fault indicator. This requires thorough fault analysis to estimate the voltage drops caused by different faults at various locations in the grid. In addition, during-fault voltage levels are dependent on the grid status at any given time. For example, the generation capacity of IBRs connected to a grid can frequently changes making fault-caused voltage drops at IBR terminals uncertain [8], [11].

Another remedy is the development of time-domain protection methods. It has been shown in [12], [13] that time-domain methods are less susceptible to IBR-caused issues. However, such methods require an advanced model with accurate parameters of the grid which is not always available in practice. In addition, it is shown that some of time-domain techniques (e.g., incremental current quantities) are affected by IBRs that jeopardizes their reliability [12].

The use of Artificial Intelligence (AI) in addressing the protection challenges arising from IBRs appears promising since AI-based schemes are faster and more reliable compared to phasor-based methods [14], [15]. However, large data sets are required for training AI-based methods making it unfavorable for power systems applications.

In [16], [17], the use of high-frequency carrier waves generated by Power Line Communication (PLC) devices is suggested for the detection of High-Impedance Faults (HIF) in medium voltage grids. However, the application of career waves in high-voltage TLs has not been addressed yet.

To address the IBR-based protection challenges, this paper proposes a new fault detection and classification method for TLs regardless of the type of power sources connected to the grid. The proposed method will reduce the protection challenges arising from IBRs and increases the hosting capacity of transmission systems for IBRs. In the proposed method, carrier waves with discernably different frequencies are transmitted from either end of a TL. Each carrier travels along one specific phase and when a fault occurs, the receiving carrier waves at the TL ends vary. Accordingly, faults are detected and classified by analyzing the variations in the receiving carriers. The main advantages of the proposed method are as follows: 1) the implementation of the proposed method is cost and time effective because it requires components similar to ones used in traditional PLC setups. 2) The response time of the proposed method is less than phasor-based relays. 3) As in the installation and commissioning process of PLCs, their outputs are tuned to be powerful enough to reach the other side of the line with a high signal-to-noise ratio, then the ambient noises, adverse weather, changes in the attenuation factor of the line, and other disturbances (e.g., switching) are not expected to affect the proposed method [18], [19], [20].

## II. Principle of the Proposed Method

The proposed method is developed based on the fact that high-frequency carrier waves (e.g., generated by PLCs) reach the other end of TLs even in adverse weather and during faults [19], [20]. However, as described in the next subsections, the characteristics of TLs change in the event of faults. Therefore, the carrier waves propagate differently in faulty TLs which leads to some changes in the receiving carriers. Therefore, the comparison of receiving carriers before and during faults enables the detection and classification of faults.

### A. Propagation of Carrier Waves Before Faults

If three carrier waves with $f_1$, $f_2$, and $f_3$ frequencies and with $v_a$, $v_b$, and $v_c$ amplitudes are injected to one side of a TL, based on the Carson's equation [21] and with respect to Fig. 1(a), the carrier voltage differences across the phases are

$$\begin{bmatrix} v_{aa'} \\ v_{bb'} \\ v_{cc'} \\ v_{gg'} \end{bmatrix} = \begin{bmatrix} z_{aa} & z_{ab} & z_{ac} & z_{ag} \\ z_{ba} & z_{bb} & z_{bc} & z_{bg} \\ z_{ca} & z_{cb} & z_{cc} & z_{cg} \\ z_{ga} & z_{gb} & z_{gc} & z_{gg} \end{bmatrix} \begin{bmatrix} i_{f_1} \\ i_{f_2} \\ i_{f_3} \\ i_g \end{bmatrix} \quad (1)$$

where $v_{aa'}$, $v_{bb'}$, and $v_{cc'}$ are carrier voltage differences across phases, $v_{gg'} = 0$ is the voltage difference between the grounds at both line ends. $z_{ij}$ is the self-impedance of phases if $i = j$, and the mutual impedance between phases $i$ and $j$ if $i \neq j$. $i_{f_1}$, $i_{f_2}$, $i_{f_3}$, and $i_g$ are carrier currents flowing in phases and the ground return path.

Even if the TL is fully transposed, neither interphase impedances nor self-impedances become the same because the carrier frequencies are different and impedances are frequency dependent. Therefore, $z_{ab} = j2\pi f_2 L_{ab}$, but $z_{ba} = j2\pi f_1 L_{ba}$ and thus $z_{ab} \neq z_{ba}$ although $L_{ab} = L_{ba}$ in fully transposed lines. The same reasoning is also applicable to the self-impedances and thus

$$\begin{cases} z_{ab} \neq z_{ba}, z_{ac} \neq z_{ca}, z_{bc} \neq z_{cb} \\ z_{aa} \neq z_{bb} \neq z_{cc} \end{cases}. \quad (2)$$

Accordingly, (1) cannot be decomposed, and each voltage difference $v_{aa'}$, $v_{bb'}$, and $v_{cc'}$ depends on all career currents $i_{f_1}$, $i_{f_2}$, $i_{f_3}$. In addition, $i_g$ contains all frequencies, and it does not become zero even in the case of fully transposed TLs ($i_g \neq 0$). Therefore, a trace of all the carrier waves exists at the receiving ends of all phase before the occurrence of faults,

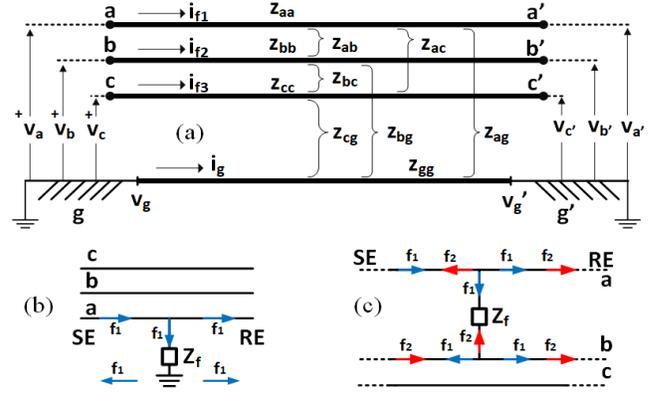

Fig. 1. (a) A typical TL with mutualities among phases and ground. (b) Wave propagation through an SLG fault between Phase A and ground. (c) Wave propagation through an LL faults between Phases A and B.

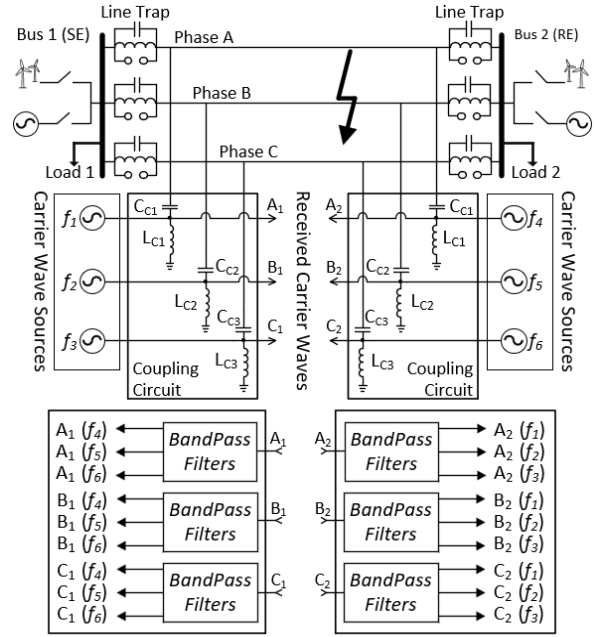

Fig. 2. The setup of the proposed method. At each phase end, a carrier wave with its own distinctive frequency is sent to the phase, and receiving carrier waves are measured through bandpass filters at the other end.

as (3) shows for Phase A.

$$v_{a'} = v_a - (z_{aa} i_{f_1} + z_{ab} i_{f_2} + z_{ac} i_{f_3} + z_{ag} i_g). \quad (3)$$

This also justifies the application of bandpass filters at the end of all the phases as shown in Fig. 2.

### B. Propagation of Carrier Waves During Faults

Faults provide additional paths for carrier waves and disturb the line model. Therefore, the amplitudes of the receiving carriers differ before and during faults. For example, Fig. 1(b) shows a Single-Line-to-Ground (SLG) fault that makes a path between Phase A and the ground. Therefore, part of the carrier enters the ground, and the receiving wave at the end of Phase A becomes distorted. In addition, Fig. 1(c) shows that the carrier traveling along Phase A (the blue arrow, $f_1$) finds its way to Phase B through the Line-to-Line (LL) fault, and then travels towards both ends of Phase B. Similarly, the carrier in Phase B (the red arrow, $f_2$) finds its way to Phase A and starts traveling towards both ends of Phase A. Thus, the receiving waves become distorted at the ends of Phases A and B.

**Algorithm 1** Logic of the Proposed Method.

If $(A_i(f_1)\&A_i(f_2)\&A_i(f_3))\downarrow \& (B_i(f_3)\&C_i(f_2))\uparrow \to$ SLG, A
If $(B_i(f_1)\&B_i(f_2)\&B_i(f_3))\downarrow \& (A_i(f_3)\&C_i(f_1))\uparrow \to$ SLG, B
If $(C_i(f_1)\&C_i(f_2)\&C_i(f_3))\downarrow \& (A_i(f_2)\&B_i(f_1))\uparrow \to$ SLG, C

If $(A_i(f_1)\&B_i(f_2))\downarrow \& (C_i(f_1)\&C_i(f_2)\&C_i(f_3))= \to$ LL, AB
If $(A_i(f_1)\&C_i(f_3))\downarrow \& (B_i(f_1)\&B_i(f_2)\&B_i(f_3))= \to$ LL, AC
If $(B_i(f_2)\&C_i(f_3))\downarrow \& (A_i(f_1)\&A_i(f_2)\&A_i(f_3))= \to$ LL, BC

If (All Amplitudes)↓ {
 If $(C_i(f_1) > A_i(f_1))\&(C_i(f_2) > A_i(f_2))\&(C_i(f_3) > A_i(f_3)) \to$ LLG, AB
 If $(B_i(f_1) > A_i(f_1))\&(B_i(f_2) > A_i(f_2))\&(B_i(f_3) > A_i(f_3)) \to$ LLG, AC
 If $(A_i(f_1) > B_i(f_1))\&(A_i(f_2) > B_i(f_2))\&(A_i(f_3) > B_i(f_3)) \to$ LLG, BC

If (All Amplitudes Are Almost Equal) → 3LG }

'=' indicates that its preceding term stays constant, '↓' means that its preceding term decreases, and '↑' means that its preceding term increases.

### C. Proposed Method Requirements and Algorithm

With reference to Fig. 2, a TL has six ends and one carrier wave with a specific frequency is injected to each end. Therefore, six different carrier waves travel along the TL. Three of them travel from Bus 1 to Bus 2, and the other three travel reversely from Bus 2 to Bus 1. The carriers are named as $P_i(f_c)$ in that $P$ indicates the phase, $i$ indicates the receiving end, and $f_c$ denotes the carrier frequency. For example, $A_2(50k)$ shows the carrier on Phase A at Bus 2 with the frequency of 50 kHz.

It is worth noting that all the carriers are expected to arrive at any given phase end due to interphase mutualities. Therefore, those are differentiated using bandpass filters as depicted in Fig. 2. Also, the impedance of transmitter and receiver units should be tunned through impedance matching process to maximize the power of carriers [18]. It is also worth clarifying that wideband line traps are installed at both ends of TLs to restrict all six carriers to the TL. Moreover, line traps limit the influencing waves generated by other components (e.g., switching actions) to enter the TLs [18], [20].

Carrier amplitudes are the most observable property of carriers requiring least signal processing. Moreover, carrier amplitudes are used in on-off modulation in PLC communications for transmitting data owing to their reliability [20]. Accordingly, the logic of the proposed method is based on carrier amplitudes and given in Algorithm 1. In that, the currently measured carrier amplitudes are continuously compared with their previous values to determine which ones increased ("↑"), decreased ("↓"), or stayed constant ("=").

As an illustration to the proposed method, Fig. 3 shows the outputs of the filters at Bus 2 in the grid shown in Fig. 2. An SLG fault occurs on Phase A at time 0.5 s and takes 0.5 s long. It is obvious in Fig. 3 that after a short transient period after the fault occurrence, all the amplitudes on Phase A decrease which is "$(A_2(f_1)\&A_2(f_2)\&A_2(f_3))\downarrow$" per the algorithm notation. At the same time, the carrier belonging to solid Phase B (i.e., 100 kHz, blue in Fig. 3), but measured on Phase C, increases. In the same way, the carrier belonging to the solid Phase C (i.e., 150 kHz, dotted black in Fig. 3), but measured on Phase B, increases. These behaviors are shown with "$(B_2(f_3)\&C_2(f_2))\uparrow$" per the notation of Algorithm 1. These changes in the carrier amplitudes satisfy the first condition in Algorithm 1 indicating that an SLG fault happened on Phase A. After an extensive number of tests, similar patterns are found for all SLG faults as shown in the first three conditions of Algorithm 1. In addition, other faults engaging different phases were tested at different locations, and the following was discovered. In the case of LL faults, all the amplitudes on the solid phase stay constant, while all other amplitudes on the faulty phases decrease. In the event of LLG and 3LG faults, all carrier amplitudes decrease, but in LLG faults, the amplitudes on the solid phase are larger than those on the faulty phases. However, as for 3LG faults, all amplitudes decrease to an almost equal level. Accordingly, Algorithm 1 was developed based on these findings.

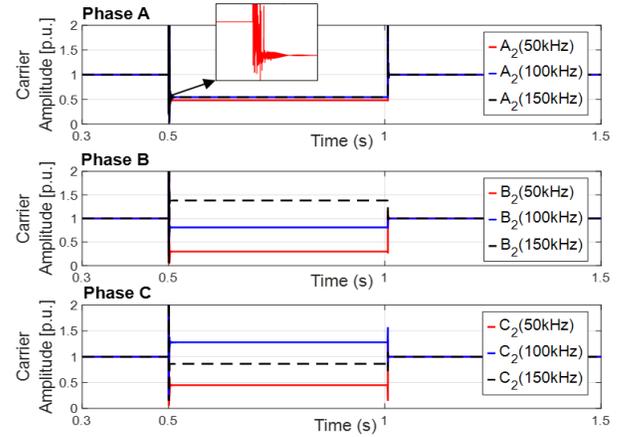

Fig. 3. Carrier waves at Bus 2 caused by an SLG fault on Phase A and 80 km away from Bus 2 and the grid is fully supplied by IBRs.

TABLE I
COMPONENTS OF COUPLING CIRCUITS AND LINE TRAPS

| Component | Carrier Frequency [kHz] | C [nF] | L [mH] |
|---|---|---|---|
| Coupling Circuit Components | $f_1 = 50$ | $C_{C1} = 20$ | $L_{C1} = 0.5$ |
| | $f_2 = 100$ | $C_{C2} = 15$ | $L_{C2} = 0.4$ |
| | $f_3 = 150$ | $C_{C3} = 10$ | $L_{C3} = 0.3$ |
| Line Trap Components | All the line traps are wideband with low and high cutoff frequencies of 40 and 310 kHz to restrict all carrier waves to the TL. | $C_T = 1.466$ [nF] $L_T = 2.709$ [mH] $R_T = 1$ [kΩ] | |

### III. TEST CASE AND RESULTS

The grid shown in Fig. 2 is used as the test system that consists of one fully transposed 100-km, 115-kV, 60-Hz TL from [22]. The Synchronous Generator's (SG) positive and zero impedances are $Z_1 = 1 + j9$ and $Z_0 = 3 + j30$ Ω. One wind farms is connected to either side of the TL. Each wind park comprises 45 doubly-fed induction generators with a capacity of 1.67 MVA. Therefore, the aggregate capacity of each park is 75 MVA. The wind park model and its parameters are found in [23]. The loads on Buses 1 and 2 are both 75 MVA. The test system can be supplied by any contribution of IBRs from 100% to zero. The frequencies of the carrier waves sent from Bus 1 to Bus 2 are $f_1 = 50$, $f_2 = 100$, and $f_3 = 150$ kHz. The parameters of the coupling circuits and line traps are selected based on [18], [20] as given in Table I. Both ends of the TL are equipped with wideband line traps with a bandwidth starting from 40 to 310 kHz in which a margin of 10 kHz is considered for both low and high cutoff frequencies to ensure the filter's desirable performance [18], [20]. In all the following experiments the fault impedance ($Z_f$) is 10 Ω, and the fault is applied to the grid when the grid is in steady state.

■ **SLG Faults:** SLG faults were applied to Phase A at different locations as given in Table II. All the carrier amplitudes (i.e.,

TABLE II
CARRIER AMPLITUDES AT BUS 2 DURING SLG FAULTS ON PHASE A.

| Received Carrier | Fault location from Bus 1 in km | | | | | | | |
|---|---|---|---|---|---|---|---|---|
| | 10 | | 20 | | 50 | | 90 | |
| | Amplitude of receiving carrier waves in p.u. | | | | | | | |
| $A_2(50k)$ | 0.41 | ↓ | 0.48 | ↓ | 0.69 | ↓ | 0.41 | ↓ |
| $A_2(100k)$ | 0.44 | ↓ | 0.51 | ↓ | 0.62 | ↓ | 0.42 | ↓ |
| $A_2(150k)$ | 0.44 | ↓ | 0.52 | ↓ | 0.62 | ↓ | 0.44 | ↓ |
| $B_2(50k)$ | 0.45 | ↓ | 0.30 | ↓ | 0.52 | ↓ | 0.41 | ↓ |
| $B_2(100k)$ | 0.84 | ↓ | 0.81 | ↓ | 0.90 | ↓ | 0.84 | ↓ |
| $B_2(150k)$ | 1.33 | ↑ | 1.38 | ↑ | 1.24 | ↑ | 1.33 | ↑ |
| $C_2(50k)$ | 0.45 | ↓ | 0.49 | ↓ | 0.48 | ↓ | 0.47 | ↓ |
| $C_2(100k)$ | 1.31 | ↑ | 1.28 | ↑ | 1.29 | ↑ | 1.31 | ↑ |
| $C_2(150k)$ | 0.85 | ↓ | 0.86 | ↓ | 0.86 | ↑ | 0.85 | ↑ |

Downward and upward arrow ("↓" and "↑") mean the carrier amplitude respectively decreases or increases from 1 p.u. to the given values during faults.

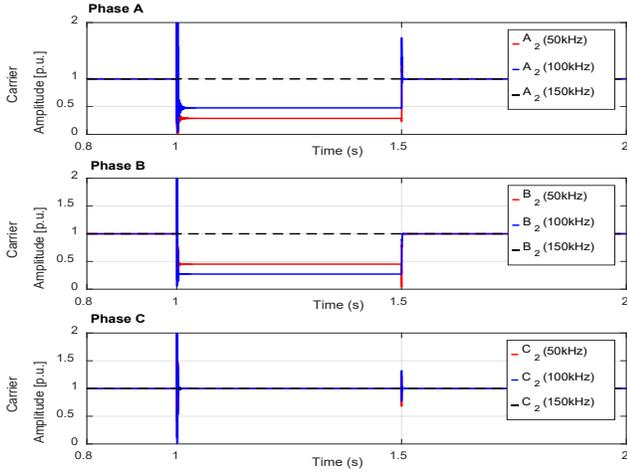

Fig. 4. Receiving carrier waves at Bus 2 before, during, and after an LL fault between Phases A and B at 20 km away from Bus 1. The system if fully supplied by IBRs. The fault occurs at t=1 s and clears at t=1.5 s.

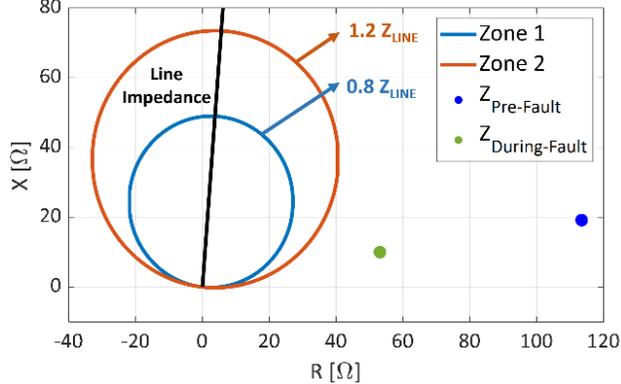

Fig. 5. The characteristic curve and performance of the distance relay.

$A_2(50k)$, $A_2(100k)$, and $A_2(150k)$) at the receiving end of Phase A drop during the fault as it is obvious in Table II. However, the receiving carriers on the other phases follow a different pattern in that the receiving carrier on Phase B from Phase C (i.e., $B_2(150k)$) and on Phase C from Phase B (i.e., $C_2(100k)$) increase as highlighted in Table II with grayed rows. The fault detection time indicates the delay between the fault inception and when the receiving carrier amplitudes become stable again and the proposed method is able to detect the fault. The longest fault detection time is 9 ms in the case of SLG faults when the system is fully IBR-based. It is noteworthy that this delay is partly due to the fluctuations in IBR outputs which mainly depends on IBR controllers' specifications.

TABLE III
CARRIER AMPLITUDES AT BUS 2 DURING LLG FAULTS ENGAGING PHASES A AND B AND THE GROUND.

| Received Carrier | Fault location from Bus 1 in km | | | | | | | |
|---|---|---|---|---|---|---|---|---|
| | 10 | | 20 | | 50 | | 90 | |
| | Amplitude of receiving carrier waves in p.u. | | | | | | | |
| $A_2(50k)$ | 0.19 | ↓ | 0.25 | ↓ | 0.47 | ↓ | 0.19 | ↓ |
| $A_2(100k)$ | 0.19 | ↓ | 0.36 | ↓ | 0.36 | ↓ | 0.19 | ↓ |
| $A_2(150k)$ | 0.57 | ↓ | 0.62 | ↓ | 0.64 | ↓ | 0.57 | ↓ |
| $B_2(50k)$ | 0.27 | ↓ | 0.29 | ↓ | 0.28 | ↓ | 0.28 | ↓ |
| $B_2(100k)$ | 0.19 | ↓ | 0.19 | ↓ | 0.25 | ↓ | 0.19 | ↓ |
| $B_2(150k)$ | 0.55 | ↓ | 0.56 | ↓ | 0.59 | ↓ | 0.55 | ↓ |
| $C_2(50k)$ | 0.73 | ↓ | 0.76 | ↓ | 0.79 | ↓ | 0.73 | ↓ |
| $C_2(100k)$ | 0.73 | ↓ | 0.76 | ↓ | 0.79 | ↓ | 0.73 | ↓ |
| $C_2(150k)$ | 0.73 | ↓ | 0.76 | ↓ | 0.79 | ↓ | 0.73 | ↓ |

Downward arrow ("↓") means a carrier amplitude decreases from 1 p.u. to the given values during faults.

TABLE IV
CARRIER AMPLITUDES AT BUS 2 DURING 3LG FAULTS.

| Received Carrier | Fault location from Bus 1 in km | | | | | | | |
|---|---|---|---|---|---|---|---|---|
| | 10 | | 20 | | 50 | | 90 | |
| | Amplitude of receiving carrier waves in p.u. | | | | | | | |
| $A_2(50k)$ | 0.04 | ↓ | 0.04 | ↓ | 0.04 | ↓ | 0.04 | ↓ |
| $A_2(100k)$ | 0.03 | ↓ | 0.03 | ↓ | 0.03 | ↓ | 0.03 | ↓ |
| $A_2(150k)$ | 0.03 | ↓ | 0.03 | ↓ | 0.03 | ↓ | 0.03 | ↓ |
| $B_2(50k)$ | 0.04 | ↓ | 0.04 | ↓ | 0.04 | ↓ | 0.03 | ↓ |
| $B_2(100k)$ | 0.04 | ↓ | 0.04 | ↓ | 0.04 | ↓ | 0.04 | ↓ |
| $B_2(150k)$ | 0.03 | ↓ | 0.03 | ↓ | 0.03 | ↓ | 0.03 | ↓ |
| $C_2(50k)$ | 0.04 | ↓ | 0.04 | ↓ | 0.04 | ↓ | 0.03 | ↓ |
| $C_2(100k)$ | 0.03 | ↓ | 0.03 | ↓ | 0.03 | ↓ | 0.03 | ↓ |
| $C_2(150k)$ | 0.04 | ↓ | 0.04 | ↓ | 0.04 | ↓ | 0.04 | ↓ |

Downward arrow ("↓") means a carrier amplitude decreases from 1 p.u. to the given values during faults.

The same tests as the ones shown in Table II were carried out on the grid, but with partial and zero IBR contribution. The behavior and amplitudes of the receiving waves stayed almost identical to the values given in Table II. This was predictable since the TL and fault properties do not depend on the supply type.

▪ LL Faults: Without loss of generality, LL faults are selected for comparing the performances of phasor-based distance relays and the proposed method because LL faults do not engage the ground. Therefore, it is not required to consider the grounding system of the grid, and the comparison results are more general. The scenario is as follows, an LL fault occurs at 20 km from Bus 1, involving Phases A and B. The received carrier waveforms are depicted in Fig. 4. It is observed that the amplitudes of the receiving carriers related to the faulty phases (i.e., $A_2(50k)$, $A_2(100k)$, $B_2(50k)$, and $B_2(100k)$) decrease. But the carrier belonging to the healthy phase C remains unchanged on the faulty phases (i.e., $A_2(150k)$, $B_2(150k)$). At the same time, all the amplitudes of receiving carriers at the end of solid Phase C stayed unchanged. The proposed method detected this fault in 8 ms.

On the other hand, a generic model of distance relays from the EMTP-RV protection library [24] was selected with cross polarization characteristics. Since there is only one TL, Zones 1 and 2 of the relay were set to 80% and 120% of the TL impedance which is a common practice in protection engineering. The positive-sequence impedances of the TL is $Z^+ = 61.376\angle 85.6^o$ Ω. Fig. 5 shows the characteristic curve of the relay together with the pre- and during-fault impedances that the relay sees. It should be noted that the shown values on

Fig. 5 are the actual values based on the primary sides of CTs and VTs for clarity purposes. It is obvious in Fig. 5 that although the fault occurs in Zone 1, the relay sees it outside of Zone 2 and remains constrained.

▪ LLG Faults: The proposed method is evaluated under LLG faults which were applied between Phases A, B, and the ground at the locations given in Table III. The values in Table III are related to the grid when it is fully supplied by IBRs. It is noted from Table III that the amplitudes of receiving carrier waves at all phases drop during LLG faults. However, all the receiving carrier amplitudes on the healthy phase (Phase C) are larger compared to ones on the faulty phases. The grayed rows of Table III show that $C_2(50k)$, $C_2(100k)$, and $C_2(150k)$ are greater than the other receiving carriers on faulty Phases A and B. The longest detection time is almost 8 ms.

▪ 3LG Faults: The performance of the proposed method under 3LG faults at the locations given in Table IV was evaluated, as well. It is observable in Table IV that all amplitudes decrease on all of the phases. Additionally, all the receiving carrier amplitudes are almost the same which is unique to 3LG faults and indicates that all the phases are affected equally. In addition, the fault detection time for 3LG faults is almost 10 ms.

## IV. Conclusion

Large-scale integration of Inverter-Based Resources (IBRs) into electric grids negatively impacts the performance of their legacy protection systems. In this paper, a source-independent method for fault detection and classification was introduced. The proposed method identifies faults based on their effects on high-frequency carrier waves traveling along Transmission Lines (TLs) and requires the components used in the setup of Power Line Communication (PLC).

The proposed method continually compares the receiving carrier waves at each end of TLs with their respective previous values. The proposed method detects the occurrence of faults based on the deviations in the amplitudes of receiving carriers. The proposed method is able to detect faults and classify their types (e.g., SLG, LL, LLG, and 3LG) within 10 ms. The types of power sources supplying the grid has no considerable effect on the performance of the proposed method as our studies on partially and fully IBR-based grids indicate. The simulation results show that the proposed method is promising to be further developed, and its sensitivity to different influencing parameters (e.g., fault impedance, arc faults, and disrupting events such as switching actions) will be evaluated.